\documentstyle[12pt]{article}

\title{{\normalsize\hfill To appear in Few Body Systems}\\*[-10pt]
       {\normalsize\hfill nucl-th/9403012}\\ ~\\
The Balian-Br\'ezin Method in Relativistic Quantum Mechanics}
\author{H.-C.~Jean$^1$,
        G.~L.~Payne$^2$ and
        W.~N.~Polyzou$^{2,}$\thanks{Research Supported by the
        U.S.~Department of Energy, Contract DE-FG02-86ER40286.}\\~\\
 $^1$Department of Physics, Florida State University\\
     Tallahassee, FL 32306, USA\\
 $^2$Department of Physics and Astronomy, The University of Iowa\\
     Iowa City, IA 52242, USA}

\begin{document}

\maketitle

\begin{abstract}
The method suggested by Balian and Br\'ezin for treating angular
momentum reduction in the Faddeev equations is shown to be applicable
to the relativistic three-body problem.
\end{abstract}

\section{Introduction}

The Faddeev equations\cite{Faddeev} provide a formulation of the
quantum mechanical three-body problem as a compact kernel integral
equation.  Compact kernel integral equations are well suited to
numerical solution because the kernel of the equation can be uniformly
approximated by a finite dimensional matrix, reducing the dynamical
equations to a system of linear algebraic equations.  Numerical
solutions of the Faddeev equations require that the abstract operator
equations be expressed as integral equations formulated in a chosen
basis.  The equations can be simplified by using bases that exploit
the rotational and translational symmetries of the problem.  The
method suggested by Balian and Br\'ezin\cite{Balian} has proved useful
in applications.  This paper illustrates how the Balian-Br\'ezin
method generalizes to the relativistic case.

\section{The Non-Relativistic Case}

The Hamiltonian for a system of three particles interacting with
two-body interactions has the form
\begin{equation}
H = H_0 + \sum_{i<j} V_{ij}~,
\label{eq:AA}
\end{equation}
where $H_0$ is the kinetic energy operator for three particles and
$V_{ij} = \hat{V}_{ij} \otimes I_k $ is the two-body interaction
between particle $i$ and $j$ imbedded in the three particle Hilbert
space.  In what follows the interaction $V_{12}$ is denoted by $V_3$,
and likewise for cyclic permutations of 1,2, and 3.

The three-body 2-cluster to 2-cluster transition operators are
\begin{equation}
T^{ij} \equiv V^j + V^i R(z) V^j~, \qquad V^i = \sum_{j\not=i} V_j~,
\label{eq:AB}
\end{equation}
where $R(z) := (z-H)^{-1}$ with $z=E \pm i0^+$.  The second resolvent
identity implies that the transition operators satisfy the coupled
integral equations:
\begin{equation}
T^{ij}(z) = V^j + \sum_{k \not= i}V_k R_k (z) T^{kj}(z)~,
\label{eq:AC}
\end{equation}
where $R_k(z) := (z-H_0-V_k)^{-1}$ is the resolvent of the Hamiltonian
for the interacting pair plus spectator.

Equations~(\ref{eq:AC}) have the same kernel as the
Alt-Grassberger-Sandhas equations\cite{Alt}, which are used instead of
the Faddeev equations in most numerical applications.

They are also representative of the type of equation to which the
Balian Br\'ezin method can be applied.  The analysis that follows is
valid for any type of connected kernel three-body equation.

In applications Eq.~(\ref{eq:AC}) is replaced by a linear system of
algebraic equations for the matrix elements of $T^{ij}(z)$ in a chosen
basis.  The complexity of the algebraic equations is reduced by
choosing a suitable basis.

The Hilbert space for the three-body system is the tensor product of
three single-particle Hilbert spaces.  Let ${\bf p}_i$, $m_i$, $s_i$
and $\mu_i$ denote the momentum, mass, spin, and magnetic quantum
number of the $i$-th particle.  Plane wave basis vectors for the
three-body system are
\begin{equation}
|{\bf p}_1 \mu_1 {\bf p}_2 \mu_2 {\bf p}_3 \mu_3 \rangle
\label{eq:AD}
\end{equation}
with normalization
\begin{equation}
\langle {\bf p}_1 \mu_1 {\bf p}_2 \mu_2 {\bf p}_3 \mu_3 | {\bf p}_1'
 \mu_1' {\bf p}_2' \mu_2' {\bf p}_3' \mu_3' \rangle =
 \prod_{i=1}^3 \delta({\bf p}_i-{\bf p}_i'\,) \delta_{\mu_i\mu_i'}~.
\label{eq:AE}
\end{equation}
The kernel of Eq.~(\ref{eq:AC})
\begin{equation}
K(i) := V_i R_i(z)
\label{eq:AF}
\end{equation}
commutes with the total linear momentum operator and the linear
momentum operator of the spectator particle, $i$.  It is separately
invariant under rotations of the spectator and the interacting pair.

The number of non-zero matrix elements of the kernel can be reduced by
evaluating it in a basis that exploits these symmetries.   One basis
commonly used is the basis of angular momentum eigenstates defined by
the following linear combination of the basis vectors (\ref{eq:AD}),
\begin{eqnarray}
| {\bf P} q_i k_i ; J \mu_J L S j l s \rangle &=& \int
 d\hat{{\bf q}}_i d\hat{{\bf k}}_i | {\bf P} {\bf q}_i {\bf k}_i ;
 \mu_i \mu_j \mu_k \rangle Y_{L\mu_L}(\hat{{\bf q}}_i)
 Y_{l\mu_l}(\hat{{\bf k}}_i)
 \langle s_j \mu_j s_k \mu_k | s \mu_s \rangle \nonumber\\
&&\times \langle l\mu_l s\mu_s| j\nu_j \rangle
 \langle s_i\mu_{i} j\nu_j | S\mu_S\rangle
 \langle L\mu_L S\mu_S | J\mu_J \rangle~,
\label{eq:AG}
\end{eqnarray}
where
\begin{eqnarray}
| {\bf P}{\bf q}_i{\bf k}_i ; \mu_1\mu_2\mu_3 \rangle &=&
 \int d{\bf p}_1 d{\bf p}_2 d{\bf p}_3
 | {\bf p}_1\mu_1 {\bf p}_2\mu_2 {\bf p}_3\mu_3 \rangle
 \delta({\bf p}_1-{\bf p}_1({\bf P},{\bf q}_i,{\bf k}_i)) \nonumber\\
&&\times
 \delta( {\bf p}_2-{\bf p}_2({\bf P},{\bf q}_i,{\bf k}_i) )
 \delta( {\bf p}_3-{\bf p}_3({\bf P},{\bf q}_i,{\bf k}_i) )
\label{eq:AH}
\end{eqnarray}
and the momenta ${\bf P}$, ${\bf q}_i$, and ${\bf k}_i$ are related to
the single particle momenta by
\begin{eqnarray}
&{\bf P} = {\bf p}_1+{\bf p}_2+{\bf p}_3~,& \\
&{\bf k}_i = {m_k\over m_{jk}}{\bf p}_j-{m_j\over m_{jk}}{\bf p}_k~,&
\label{eq:XA}
\end{eqnarray}
and
\begin{equation}
{\bf q}_i = {m_{jk}\over M}{\bf p}_i -{m_i\over M}{\bf p}_{jk}~,
\label{eq:AI}
\end{equation}
where ${\bf p}_{jk} := {\bf p}_j + {\bf p}_k$, $M=m_1+m_2+m_3$,
$m_{jk}=m_j+m_k$, and $i,j,k$ are cyclic permutations of $1,2,3$.
With this choice the Jacobian of the variable change
$\{ {\bf P},{\bf q}_i,{\bf k}_i \} \to
\{ {\bf p}_1,{\bf p}_2,{\bf p}_3 \}$
is unity.

The matrix elements of the kernel of Eq.~(\ref{eq:AC}) in this basis
are
\begin{eqnarray}
\lefteqn{\langle {\bf P} q_i k_i ; J \mu_J L S j l s | V_i R_i (z) |
 {\bf P}' q_i' k_i' ; J' \mu_J' L' S' j' l' s' \rangle} \nonumber\\
&=&\delta({\bf P}-{\bf P}') {\delta(q_i-q_i')\over q_i^2} \delta_{JJ'}
 \delta_{\mu_J\mu_J'} \delta_{jj'} \delta_{LL'}\delta_{SS'}\nonumber\\
&&\times\langle k_i j l s | \hat{V}_i \hat{R}_i \left( z-
 {{P}^2\over 2M}-q^2_i{M\over2m_im_{jk}}\right) | k_i'jl's'\rangle~,
\label{eq:AJ}
\end{eqnarray}
where
$$ \hat{V}_i \hat{R}_i \left( z - {{P}^2\over 2M} - q^2_i
 {M\over 2 m_i m_{jk}} \right) $$
is
\begin{equation}
\hat{V}_{i} \left[z - {{P}^2\over 2M} - q^2_i {M\over 2 m_i m_{jk} }
 - k_i^2 {m_{jk} \over 2 m_j m_k } - \hat{V}_i \right]^{-1}
\label{eq:AK}
\end{equation}
and $\hat{V}_i$ is the two-body interaction.

The basis~(\ref{eq:AG}) exploits the translational and rotational
invariance of the system and the spectator particle.  The coupling
implicit in Eq.~(\ref{eq:AC}) breaks the translational and rotational
invariance associated with the original spectator particle.  This can
be illustrated by considering the iterated kernel.  Let $K(i)$ denote
the kernel with particle $i$ as spectator and let $|i\rangle$ denote
the basis vector~(\ref{eq:AG}) corresponding to particle $i$ being the
spectator.  The matrix elements of the iterated kernel are
\begin{equation}
\sum_{j\not=i} \langle i | K(i)K(j) | j \rangle = \sum_{j\not=i}
 \langle i | K(i) | i \rangle \langle i | j \rangle
 \langle j | K(j) | j \rangle~,
\label{eq:AL}
\end{equation}
where in order to express the kernel in a basis where it has the
form~(\ref{eq:AJ}) it is necessary to introduce the change of basis
$\langle i|j\rangle$.  Since $j\not=i$ this changes the spectator and
breaks the invariance associated with the spectator particle.

The change of basis is computed from the definitions by transforming
the bras and kets in Eq.~(\ref{eq:AJ}) to the tensor product of
one-body basis vectors and evaluating the overlap.  The general result
is determined by taking cyclic permutations of
\begin{eqnarray}
\langle 1 | 2' \rangle &=& \delta({\bf P}-{\bf P}') \int
 d\hat{{\bf q}}_1 d\hat{{\bf k}}_1 d\hat{{\bf q}}_2 d\hat{{\bf k}}_2
 \delta( {\bf q}_1 - {\bf q}_1({\bf q}_2,{\bf k}_2 ) )
 \delta( {\bf k}_1 - {\bf k}_1({\bf q}_2,{\bf k}_2 ) ) \nonumber\\
&&\times
 Y^*_{L \mu_{L}}(\hat{{\bf q}}_1)
 Y^*_{l\mu_{l}}(\hat{{\bf k}}_1)
 Y _{L' \mu_{L}'}( \hat{{\bf q}}_2 )
 Y_{l'\mu_{l}'}(\hat{{\bf k}}_2) \nonumber\\
&&\times
 \langle J \mu_J | L \mu_L S \mu_S \rangle
 \langle S \mu_S | s_1 \mu_{1} j \nu_j \rangle
 \langle j \nu_j | l \mu_l s \mu_s \rangle
 \langle s \mu_s | s_2 \mu_2 s_3 \mu_3 \rangle \nonumber\\
&&\times
 \langle s_3 \mu_3 s_1 \mu_1 | s' \mu_s' \rangle
 \langle l' \mu_l' s' \mu_s'| j' \nu_j' \rangle
 \langle s_2 \mu_2 j' \nu_j' | S' \mu_S' \rangle
 \langle L' \mu_L' S' \mu_S' | J' \mu_J' \rangle~,
\label{eq:AM}
\end{eqnarray}
where the sums over all repeated magnetic quantum numbers are
implicit.  This expression includes four two-dimensional integrals
over the angles associated with each relative momenta.  Two of the
integrals can be done in terms of the angular parts of the delta
functions.  This still leaves two two-dimensional integrals over the
angles and two delta functions that fix the magnitude of the relative
momenta.  In this form the change of basis is a complicated
transformation to implement numerically.

Balian and Br\'ezin exploit the rotational invariance of the overlap
matrix elements~(\ref{eq:AM}) to facilitate the computation.
Rotational invariance implies that the matrix element~(\ref{eq:AM}) is
equal to kronecker deltas in $\delta_{JJ'}\delta_{\mu_J\mu_{J'}}$
multiplied by an expression independent of $\mu_J$.  It follows
that~(\ref{eq:AM}) is equal to its average over $\mu_J$.  The
averaging makes the integrand in~(\ref{eq:AM}) invariant under
simultaneous rotations of the vectors ${\bf q}_1$, ${\bf k}_1$,
${\bf q}_2$, and ${\bf k}_2$ which implies that the integrand depends
only on the independent invariants $q_1^2$, $k_1^2$ and
${\bf q}_1\cdot{\bf k}_1$.  The quantity ${\bf q}_1\cdot{\bf k}_1$ is
fixed in terms of $q_1^2$, $q_2^2$, and $k_2^2$ or $k_1^2$, $q_2^2$,
and $k_2^2$.  The result is that, after the delta function fixes the
integral over the cosine of the angle between $\hat{{\bf q}}_1$ and
$\hat{{\bf k}}_1$, the integral over the remaining angles is equal to
a phase-space factor times the invariant integrand.  The integrand can
be evaluated by choosing the vectors ${\bf q}_1$, ${\bf k}_1$,
${\bf q}_2$, and ${\bf k}_2$ to have any convenient values consistent
with the kinematics:
\begin{eqnarray}
\langle 1 | 2' \rangle &=& \delta({\bf P}-{\bf P}') \delta(E-E')
 {\delta_{J J'}\delta_{\mu_J \mu_J'} \over 2J+1} \nonumber\\
&&\times { 8 \pi^2 m_{23} m_{13} \over m_1 m_2 m_3 k_1 k_2 q_1 q_2}
 Y^* _{L \mu_{L}}( \hat{{\bf q}}_1 )
 Y^*_{l\mu_{l}}(\hat{{\bf k}}_1)
 Y _{L' \mu_{L}'}( \hat{{\bf q}}_2 )
 Y_{l'\mu_{l}'}(\hat{{\bf k}}_2) \nonumber\\
&&\times
 \langle J \nu | L \mu_L S \mu_S \rangle
 \langle S \mu_S | s_1 \mu_{1} j \mu_j \rangle
 \langle j \mu_j | l \mu_l s \mu_s \rangle
 \langle s \mu_s | s_2 \mu_2 s_3 \mu_3 \rangle \nonumber\\
&&\times
 \langle s_3 \mu_3 s_1 \mu_1 | s' \mu_s' \rangle
 \langle l' \mu_l' s' \mu_s'| j' \mu_j' \rangle
 \langle s_2 \mu_2 j' \mu_j' | S' \mu_S' \rangle
 \langle L' \mu_L' S' \mu_S' | J \nu \rangle~,
\label{eq:AN}
\end{eqnarray}
where $E$ is the kinetic energy and the unit vectors are fixed by
${\bf k}_i$ and ${\bf q}_i$, which can be chosen arbitrarily subject
to the constraints that fix $q_i^2$, $k_i^2$, and
${\bf k}_i\cdot{\bf q}_i$.  Note that the above expression includes a
sum over the overall magnetic quantum number, $\nu$.

Balian and Br\'ezin suggest three choices of fixing the direction of
${\bf k}_i$ and ${\bf q}_i$, subject to the constraint on the scalar
product, which facilitate the computation of these expressions.  These
three choices lead to additional simplifications.  This method for
treating the change of basis matrix elements is employed in many
existing numerical treatments of the three-body problem.  The same
considerations apply to configuration space computations.

The form of the recoupling coefficient in Eq.~(\ref{eq:AN}) uses one
of the delta functions in the two momentum variables to do the
integral over $\hat{{\bf k}}_i\cdot\hat{{\bf q}}_i$ while the other
gives the energy-conserving delta functions.  In principle these delta
functions can be used to perform integrals over any pair of variables.
In applications it is convenient to use the delta functions in the
relative momenta to do the two integrals over the relative momenta.
In this case the one-dimensional integral over
$\hat{{\bf k}}_i\cdot\hat{{\bf q}}_i$ must be calculated numerically.

\section{The Relativistic Case}

The Balian-Br\'ezin method is also applicable to large class of
relativistic formulations of the three-body problem, which includes
all generalized Bakamjian-Thomas\cite{Bak} formulations, including
specific Bakamjian-Thomas formulations in any of Dirac's\cite{Dirac}
forms of the dynamics.

Relativistic invariance of a quantum model requires that all
probabilities have values independent of the choice of inertial
coordinate system, where by definition any two inertial coordinate
systems are related by Poincar\'e transformations continuously
connected to the identity.  Wigner's theorem states that this
condition implies the existence\cite{Wigner} of a unitary ray
representation of the Poincar\'e group on the three-body Hilbert
space.

The problem of constructing the unitary representation of the
Poincar\'e group plays the same role in relativistic quantum mechanics
as constructing the unitary representation of the one-parameter time
evolution group in non-relativistic quantum mechanics.  In the
non-relativistic case it is sufficient to solve the eigenvalue problem
for the Hamiltonian.  The one-parameter time evolution group is
expressed in terms of the eigenfunctions and eigenvalues of the
Hamiltonian as
\begin{equation}
U(t) = \sum_E | E \rangle e^{-iEt} \langle E |~.
\label{eq:AO}
\end{equation}
In the relativistic case it is sufficient to solve the eigenvalue
problem for the Casimir operators of the Poincar\'e group.  For system
of massive particles this is equivalent to finding the simultaneous
eigenstates of the mass and spin operators.  Representation theory for
the Poincar\'e group is used to construct the unitary representation,
$U(\Lambda,a)$, of the Poincar\'e group in terms of the eigenvalues
and eigenstates of the mass and spin similar to the manner that the
eigenvalues and eigenstates of $H$ are used to construct $U(t)$ in
Eq.~(\ref{eq:AO}).

Bakamjian-Thomas models are models with interactions that commute with
and are independent of three independent functions of the
non-interacting four momentum and all components of a chosen
non-interacting spin.  The analysis that follows is limited to
Bakamjian-Thomas type models.  The dynamical problem in a
Bakamjian-Thomas dynamics is to find simultaneous eigenstates of the
mass and non-interacting spin operator in a suitable basis.  The
Balian-Br\'ezin method can be applied to Bakamjian-Thomas models
because the mass operator commutes with the non-interacting spin
vector, however, the explicit justification in the relativistic case
is more complicated because the quantities that replace ${\bf q}_i$
and ${\bf k}_i$ in the relativistic case do not generally transform as
vectors under rotations.

The mass operator in the relativistic case has the same form as the
Hamiltonian in the non-relativistic case, consisting of a mass
operator $M_0$ for three non-interacting particles plus pairwise
interactions
\begin{equation}
M = M_0 + \sum_{i < j} V_{ij}~.
\label{eq:AP}
\end{equation}
Time dependent scattering theory \cite{Keister} gives the transition
operators in the relativistic case,
\begin{equation}
T^{ij} : = V^j + V^i R(z) V^j~, \qquad V^i = \sum_{j\not=i} V_j~,
\label{eq:AQ}
\end{equation}
where $R(z)=(z-M)^{-1}$ is the resolvent of the mass operator.  The
second resolvent identity implies the components of the transition
operators satisfy the coupled equations
\begin{equation}
T^{ij}(z) = V^j + \sum_{k \not= i}V_k R_k (z) T^{kj}(z),
\label{eq:AR}
\end{equation}
where for the relativistic equation $R_k(z):=(z-M_0-V_k)^{-1}$ is the
resolvent of the mass operator for the interacting pair plus
spectator.  Eq.~(\ref{eq:AR}) has the same form as the
non-relativistic equation.  Differences appear in the relation of the
two-body interactions in the three-particle Hilbert space and the
two-body interactions in the two-body Hilbert space.  The structure of
the interactions is given in Eq.~(\ref{eq:BO}) below.

The Hilbert space for a single particle in relativistic quantum
mechanics is an irreducible representation space for the Poincar\'e
group corresponding to the mass and spin of the particle.  Vectors in
this space can be expanded as linear combinations of simultaneous
eigenstates of the linear momentum and $z$-component of spin.  In
relativistic quantum mechanics there are many different spin operators
that satisfy $SU(2)$ commutation relations whose square is the total
spin.  Although the spectrum of the magnetic quantum numbers is fixed
by the $SU(2)$ commutation relations, the physical interpretation of
the spin operator is determined by the transformation properties of
the single particle states under the Poincar\'e group.  In general the
single particle vectors $|{\bf p}\mu\rangle$ transform as mass-$m$
spin-$s$ irreducible representations of the Poincar\'e group
\begin{equation}
U(\Lambda,a) | {\bf p} \mu \rangle = e^{i \Lambda p_m \cdot a }
 | {\bf p}_\Lambda  \mu' \rangle
 \Biggl| {\omega_m({\bf p}_\Lambda) \over \omega_m({\bf p})}
 \Biggr|^{1/2} D^s_{\mu'\mu} (B^{-1}(\Lambda p_m) \Lambda B(p_m))~,
\label{eq:AS}
\end{equation}
where $\omega_m({\bf p}):=\sqrt{{\bf p}^2+m^2}$ is the energy,
$p_m=(\omega_m({\bf p}),{\bf p})$ is the four momentum of a particle
with mass $m$ and momentum ${\bf p}$, $p_\Lambda:=\Lambda p_m$ is the
Lorentz transform of the four momentum $p_m$, and $B(p_m)$ is a
Lorentz-boost-valued function of $p_m$ with the property
\begin{equation}
B(p_m)p_0 = p_m~,
\label{eq:AT}
\end{equation}
where $p_0:=(m,0,0,0)$ is the 0-momentum four-vector.  The quantity
\begin{equation}
\Biggl|{\omega_m({\bf p}_\Lambda)\over\omega_m({\bf p})}\Biggr|^{1/2}
 =\Biggl|{\partial{\bf p}_\Lambda\over\partial{\bf p}}\Biggr|^{1/2}
\label{eq:AU}
\end{equation}
fixes the normalization of the transformed state to ensure the
unitarity of $U(\Lambda,a)$.

The combination $R_w(\Lambda,p_m):=B^{-1}(\Lambda p_m)\Lambda B(p_m)$
is a rotation for any $\Lambda$, called the Wigner rotation associated
with the boost $B(p_m)$. The interpretation of the magnetic quantum
number is determined by the choice of $B(p_m)$.  A boost $B(p_m)$ is
defined up to a rotation valued function $R(p_m)$ of $p_m$
\begin{equation}
B(p_m) \to B'(p_m) = B(p_m) R(p_m)~.
\label{eq:AV}
\end{equation}
The canonical boost $B_c(p_m)$ is the unique boost with the properties
\begin{equation}
B_c(p_0) = I
\label{eq:AW}
\end{equation}
and
\begin{equation}
R_w(R,p_m) = B_c^{-1}(R p_m) R B_c(p_m) = R
\label{eq:AX}
\end{equation}
for any rotation $R$.  Eq.~(\ref{eq:AX}) states that for canonical
boosts the Wigner rotation of a rotation is the rotation itself.  This
is not true for other types of boosts.

Any other boost is related to a canonical boost by a $p$-dependent
rotation as in Eq.~(\ref{eq:AV}).  This rotation is called a
generalized Melosh rotation\cite{Keister}.  The helicity spin and
front-form spin are examples of spins corresponding to non-canonical
boosts\cite{Keister}.  These two choices are distinguished by other
special properties.

A basis for the three-particle Hilbert space is the tensor product of
three single particle bases.  Basis vectors have the form
\begin{equation}
| {\bf p}_1 \mu_1 {\bf p}_2 \mu_2 {\bf p}_3 \mu_3 \rangle
\label{eq:AY}
\end{equation}
with the same normalization convention as the non-relativistic
expression~(\ref{eq:AD}).

A basis for the three-body system that simplifies the matrix elements
of the kernel $K(i)=V_i R_i(z)$ of the relativistic three-body
equations~(\ref{eq:AR}) is constructed by finding the coefficients of
the linear transformation that take the product of three irreducible
representations of the Poincar\'e group to a direct integral of
irreducible representations.  This is equivalent to the problem of
constructing Clebsch-Gordan coefficients of the Poincar\'e group.
This replaces the successive pairwise coupling of the irreducible
representation spaces of the Euclidean group (rotations and
translations) used in the non-relativistic reduction.

The construction of the Clebsch-Gordan coefficients of the Poincar\'e
group is most easily understood by successive coupling of pairs of
irreducible representations.  To treat the general case first consider
the case of canonical spin.  Kinematic variables for the two particle
system are
\begin{equation}
P=p_1+p_2~,\qquad m_{12}=\sqrt{-P\cdot P}~,\qquad k_i=B_c^{-1}(P)p_i~,
\label{eq:AZ}
\end{equation}
corresponding to the total four momentum of the non-interacting pair,
the invariant mass of the non-interacting pair, and the relative
momentum of the non-interacting pair, respectively.  The relative
momentum defined in Eq.~(\ref{eq:AZ}) is not a true four-vector.  It
undergoes Wigner rotations when the system is Lorentz-transformed.

The two-body basis vectors that transform irreducibly under the action
of the tensor product of one-body representations of the Poincar\'e
group are
\begin{eqnarray}
| {\bf P} \mu (k j l s) \rangle &:=& U_1\bigl[B_c(P)\bigr]\otimes
 U_2\bigl[B_c(P)\bigr] \int d\hat{{\bf k}} | {\bf k}, \mu_1, -{\bf k},
 \mu_2 \rangle Y_{l\mu_l}(\hat{{\bf k}}) \nonumber\\
&&\times\langle s_1 \mu_1 s_2 \mu_2 | s \mu_s \rangle
 \langle l \mu_l s \mu_s | j \mu \rangle \nonumber\\
&=&\int d\hat{{\bf k}} | {\bf p}_1({\bf P},{\bf k}), \mu_1',
 {\bf p}_2({\bf P},{\bf k}), \mu_2' \rangle
 \Biggl| {\omega_{m_1}({\bf p}_1) \over \omega_{m_1}({\bf k}_1)}
 {\omega_{m_2}({\bf p}_2) \over \omega_{m_2}({\bf k}_2)}
 {m_{12} \over \omega_{m_{12}}({\bf P})} \Biggr|^{1/2} \nonumber\\
&&\times D^{s_1}_{\mu_1'\mu_1}[R_{wc}(B_c (P),k_1)]
 D^{s_2}_{\mu_2'\mu_2}[R_{wc}(B_c (P),k_2)]
 Y_{l \mu_l}(\hat{{\bf k}}_1) \nonumber\\
&&\times\langle s_1 \mu_1 s_2 \mu_2 | s \mu_s \rangle
 \langle l \mu_l s \mu_s | j \mu \rangle~,
\label{eq:BA}
\end{eqnarray}
where ${\bf k}={\bf k}_1=-{\bf k}_2$.  The factor
$m_{12}/\omega_{m_{12}}({\bf P})$ fixes the normalization to be
\begin{equation}
\langle {\bf P} \mu (k j l s) | {\bf P}' \mu' (k' j' l' s') \rangle
 = \delta({\bf P}-{\bf P}') {\delta(k-k')\over k^2} \delta_{jj'}
 \delta_{\mu\mu'} \delta_{ll'} \delta_{ss'}~.
\label{eq:BB}
\end{equation}
The single particle and combined spins can be transformed from
canonical to any other type of spin (i.e., a spin associated with an
arbitrary boost $B_x(p_m)$) with the unitary
transformation\cite{Keister}
\begin{equation}
| {\bf p} \mu \rangle_x = | {\bf p} \nu \rangle_c
 D^j_{\nu\mu}[R_{cx}({\bf p})]~, \qquad
 R_{cx}({\bf p}) := B_c^{-1}({\bf p}) B_x({\bf p})~,
\label{eq:BC}
\end{equation}
where $R_{cx}({\bf p})$ is a generalized Melosh rotation.  A direct
calculation using the relations
\begin{equation}
Y_{l\mu_l}(\hat{{\bf k}}_c) D^l_{\mu_l\nu_l}[R_{xc}({\bf p})] =
 Y_{l\nu_l}(R_{xc}({\bf p})\hat{{\bf k}}_c) =
 Y_{l\nu_l}(\hat{{\bf k}}_x)
\label{eq:BD}
\end{equation}
and
\begin{equation}
R_{xc}({\bf p}_1) R_{wc}(B_c(p),k_{1c}) R_{cx}({\bf p}) =
 R_{wx}(B_x(p),k_{1x}) R_{xc}({\bf k}_{1x}) \qquad (1\leftrightarrow2)
\label{eq:BE}
\end{equation}
shows
\begin{eqnarray}
| {\bf P} \mu (k j l s) \rangle_x &:=& \int d\hat{{\bf k}}
 | {\bf p}_1({\bf P},{\bf k}), \mu_1', {\bf p}_2({\bf P},{\bf k}),
 \mu_2' \rangle_x ~
 \Biggl| {\omega_{m_1}({\bf p}_1)\over\omega_{m_1}({\bf k}_1)}
 {\omega_{m_2}({\bf p}_2)\over\omega_{m_2}({\bf k}_2)}
 {m_{12}\over\omega_{m_{12}}({\bf P})} \Biggr|^{1/2} \nonumber\\
&&\times D^{s_1}_{\mu_1'\mu_1}[R_{wx}(B_x(P),k_{1x})
 R_{xc}({\bf k}_{1x})] D^{s_2}_{\mu_2'\mu_2}[R_{wx}(B_x(P),k_{2x})
 R_{xc}({\bf k}_{2x})] \nonumber\\
&&\times Y_{l\mu_l}(\hat{{\bf k}}_{1x}) \langle s_1 \mu_1 s_2 \mu_2
 | s \mu_s \rangle \langle l \mu_l s \mu_s | j \mu \rangle~.
\label{eq:BF}
\end{eqnarray}
In expression~(\ref{eq:BF}) the argument of the $D$-functions is the
product of a generalized Melosh rotation followed by the Wigner
rotation associated with the $x$-spin.  The quantity ${\bf k}_x$
is related to the canonical ${\bf k}_c$ by a generalized Melosh
rotation
\begin{equation}
{\bf k}_x = B_x^{-1}({\bf p}) B_c({\bf p}) {\bf k}_c =
 R_{xc}({\bf p}) {\bf k}_c~.
\label{eq:BG}
\end{equation}
As a result of this construction the state
$|{\bf P}\mu(kjls)\rangle_x$ transforms like a particle with mass
$m_{ij}=\sqrt{m_i^2+{\bf k}^2}+\sqrt{m_j^2+{\bf k}^2}$ and spin $j$
where the magnetic quantum number transforms as an $x$-spin:
\begin{eqnarray}
\lefteqn{U_i(\Lambda,a) \otimes U_j(\Lambda,a)
 | {\bf P} \mu (k j l s) \rangle_x} \nonumber\\
&&= e^{i\Lambda P\cdot a} | {\bf P}_\Lambda \mu' (k j l s) \rangle_x ~
 \Biggl| {\omega_{m_{ij}}({\bf P}_\Lambda) \over \omega_{m_{ij}}
 ({\bf P})} \Biggr|^{1/2} D^j_{\mu'\mu}[R_{wx}(\Lambda,P)]~.
\label{eq:BH}
\end{eqnarray}
The transformation properties of this two-particle state are identical
to those of a single particle.

To construct the three particle basis the tensor product for a pair
plus spectator
\begin{equation}
| {\bf p}_{jk} \mu_{jk} (k_{i} j l s) \rangle_x \otimes
 | {\bf p}_i \mu_i \rangle_x
\label{eq:BI}
\end{equation}
is decomposed into a direct integral of irreducible representations by
repeating the above analysis replacing one of the single particle
states by the two-body state~(\ref{eq:BF}).  The new kinematic
quantities are
\begin{equation}
P=p_{i}+p_{jk}~, \qquad M_0=\sqrt{- P\cdot P}~, \qquad
 q_i=B_c^{-1}(P)p_i~, \qquad  q_{jk}=B_c^{-1}(P)p_{jk}~,
\label{eq:BJ}
\end{equation}
where ${\bf q}:={\bf q}_i=-{\bf q}_{jk}$.  In addition, the two-body
kinematic variables are
\begin{equation}
k_j = B_c^{-1}(p_{jk})p_j, \qquad
k_k = B_c^{-1}(p_{jk})p_k, \qquad {\bf k}:= {\bf k}_j =-{\bf k}_k~.
\label{eq:BK}
\end{equation}
The resulting basis is related to the single particle bases by
\begin{eqnarray}
| {\bf P}\mu;qJLSkjls \rangle &:=& \int d\hat{{\bf q}} d\hat{{\bf k}}
 | {\bf p}_i\mu_i' {\bf p}_j\mu_j' {\bf p}_k\mu_k' \rangle \nonumber\\
&&\times\Biggl| {\omega_{m_j}({\bf p}_j)\over\omega_{m_j}({\bf k}_j)}
 {\omega_{m_k}({\bf p}_k)\over\omega_{m_k}({\bf k}_k)}
 {m_{jk}\over\omega_{m_{jk}}({\bf p}_{jk})}
 {\omega_{m_i}({\bf p}_i)\over\omega_{m_i}({\bf q}_i)}
 {\omega_{m_{jk}}({\bf p}_{jk})\over\omega_{m_{jk}}({\bf q}_{jk})}
 {M_0\over\omega_{M_0}({\bf P})} \Biggr|^{1/2} \nonumber\\
&&\times D^{s_j}_{\mu_j'\mu_j}[R_{wx}(B_x(p_{jk}),k_{jx})R_{xc}(k_j)]
 D^{s_k}_{\mu_k'\mu_k}[R_{wx}(B_x(p_{jk}),k_{kx})R_{xc}(k_{kx})]
 \nonumber\\
&&\times Y_{l\mu_l}(\hat{{\bf k}}_{ix}) \langle s_j \mu_j s_k \mu_k |
 s \mu_s \rangle \langle l \mu_l s \mu_s | j\nu_j' \rangle \nonumber\\
&&\times D^{j}_{\nu_j' \nu_j}[R_{wx}(B_x(P),q_{jkx})R_{xc}(q_{jkx})]
 D^{s_i}_{\mu_i'\mu_i}[R_{wx}(B_x(P),q_{ix})R_{xc}(q_{ix})]\nonumber\\
&&\times Y_{L\mu_L}(\hat{{\bf q}}_{ix}) \langle j \nu_j s_i \mu_i |
 S \mu_S \rangle \langle L \mu_L S \mu_S | J \mu_J \rangle~.
\label{eq:BL}
\end{eqnarray}
This basis also transforms irreducibly under the tensor product of
three one-body representation of the Poincar\'e group,
\begin{eqnarray}
\lefteqn{U_1(\Lambda,a) \otimes U_2(\Lambda,a) \otimes U_3(\Lambda,a)
 | {\bf P} \mu (q J L S k j l s) \rangle_x} \nonumber\\
&&=e^{i P_\Lambda\cdot a} | {\bf P}_\Lambda\mu'(qJLSkjls) \rangle_x
 \Biggl| {\omega_{M_0}({\bf P}_\Lambda)\over\omega_{M_{0}}({\bf P})}
 \Biggr|^{1/2} D^J_{\mu'\mu}[R_{wx}(\Lambda,P)]~.
\label{eq:BM}
\end{eqnarray}
The unitary representation of the Poincar\'e group in
Eq.~(\ref{eq:BM}) is not the physical representation.  What is
relevant is that in a Bakamjian-Thomas formulation both the
interacting and non-interacting representation of the Poincar\'e group
share the same spin operator.

There is a collection of Bakamjian-Thomas models associated with each
different type of irreducible basis element of the form~(\ref{eq:BL})
that are unitarily equivalent and have the same scattering matrix
elements.  The kernel of the integral equation is defined in terms of
its matrix elements in the appropriate basis
\begin{eqnarray}
\lefteqn{\langle {\bf P} \mu ; q J L S k j l s | V_i R_i (z) |
 {\bf P}' \mu' ; q' J' L' S' k' j' l' s' \rangle}\nonumber\\
&& = \delta({\bf P}-{\bf P}') \delta_{JJ'} \delta_{\mu\mu'}
 {\delta(q-q')\over q^2} \delta_{LL'} \delta_{SS'} \delta_{jj'}
 \langle kjls | \hat{V}_i (q^2) \hat{R}_i(z;q^2) | k'jl's' \rangle~,
\label{eq:BN}
\end{eqnarray}
where the operators $\hat{V}_i(q^2)\hat{R}_i(z;q^2)$ are related to
two-body interactions $v_i$ with matrix elements
$\langle jkls|v_i|jk'l's'\rangle$ by
\begin{eqnarray}
\hat{V}_i(q^2) \hat{R}_i(z;q^2) &=& \Bigl[ \sqrt{{\bf q}^2+(m_{jk}+
 v_{i})^2} - \sqrt{{\bf q}^2+m^2_{jk}} \, \Bigr] \nonumber\\
&&\times \Bigl[ z - \sqrt{{\bf q}^2+m_i^2} - \sqrt{{\bf q}^2+
 (m_{jk}+v_{i})^2} \,\, \Bigr]^{-1}~.
\label{eq:BO}
\end{eqnarray}
As in the non-relativistic case, this can be computed only after
solving the two-body problem.  What is important is the collection of
delta functions on the right hand side of Eq.~(\ref{eq:BN}).  For
Bakamjian-Thomas models, where the functions of the four-momentum that
commute with the interaction are not the three components of the
linear momentum, the delta functions in the components of the linear
momentum must be replaced by delta functions in the appropriate
functions of the non-interacting four-momentum.  For instance, in
Dirac's point-form dynamics\cite{Dirac} the three components of the
four-velocity replace the three components of the momentum.  In the
case of Dirac's front-form dynamics\cite{Dirac} the delta functions in
the components of the three momentum are replaced by delta functions
in the three components of the four momentum that generate
translations tangent to the light front, $x^3+t=0$.  The front-form
case also requires a special choice of spin operator.

As in the non-relativistic case, when Eq.~(\ref{eq:AR}) is iterated, a
new interaction is introduced, violating the symmetries associated
with the initial spectator particle.  Thus if
$\langle i|K(i)|i\rangle$ denotes the expression in Eq.~(\ref{eq:BN})
the matrix elements of the iterated kernel are
\begin{equation}
\sum_{j\not=i} \langle i | K(i)K(j) | j \rangle =
\sum_{j\not=i} \langle i | K(i) | i' \rangle \langle i' | j' \rangle
 \langle j' | K(j) | j \rangle~.
\label{eq:BP}
\end{equation}
In order to take advantage of the simple form of the kernel in
Eq.~(\ref{eq:BN}) it is necessary to compute the overlap
$\langle i|j'\rangle$ where
\begin{equation}
\langle1|2'\rangle := \langle{\bf P}\mu; q_1 J L_1 S_1 k_1 j_1 l_1 s_1
 | {\bf P}' \mu' ; q_2' J' L_2' S_2' k_2' j_2' l_2' s_2' \rangle~,
\label{eq:BQ}
\end{equation}
and cyclic permutations.  By direct calculation the matrix element is
\begin{eqnarray}
\langle 1 | 2' \rangle &=& \int d\hat{{\bf q}}_1  d\hat{{\bf k}}_1
 \int d\hat{{\bf q}}'_2 d\hat{{\bf k}}'_2 \prod_{i=1}^3
 \delta[{\bf p}_i({\bf P},{\bf q}_1,{\bf k}_1) -
 {\bf p}_i({\bf P}',{\bf q}'_2,{\bf k}'_2)] \nonumber\\
&& \times \Biggl|
 {\omega_{m_2}({\bf p}_2)\over\omega_{m_2}({\bf k}_{2x})}
 {\omega_{m_3}({\bf p}_3)\over\omega_{m_3}({\bf k}_{3x})}
 {m_{23}\over\omega_{m_{23}}({\bf p}_{23})}
 {\omega_{m_1}({\bf p}_1)\over\omega_{m_1}({\bf q}_{1x})}
 {\omega_{m_{23}}({\bf p}_{23})\over\omega_{m_{23}}({\bf q}_{23x})}
 {M_0\over\omega_{M_0}({\bf P})} \nonumber\\
&&\times
 {\omega_{m_3}({\bf p}_3')\over\omega_{m_3}({\bf k}_{3x}')}
 {\omega_{m_1}({\bf p}_1')\over\omega_{m_1}({\bf k}'_{1x})}
 {m_{31}'\over\omega_{m_{31}'}({\bf p}'_{31})}
 {\omega_{m_2}({\bf p}'_2)\over\omega_{m_2}({\bf q}'_{2x})}
{\omega_{m'_{31}}({\bf p}'_{31})\over\omega_{m_{31}'}({\bf q}_{31x}')}
 {M_0'\over\omega_{M_0'}({\bf P}')} \Biggr|^{1/2} \nonumber\\
&&\times \langle J \mu | L \mu_L S \mu_S \rangle \langle S \mu_S |
 j \mu_j s_1 \mu_1 \rangle Y_{L\mu_L}^*(\hat{\bf q}_{1x}) \nonumber\\
&&\times D^{j}_{\mu_j\nu_j}[R_{cx}(q_{23x})R_{wx}(B_x^{-1}(P),p_{23})]
 D^{s_1}_{\mu_1\nu_1}[R_{cx}(q_{1x})R_{wx}(B_x^{-1}(P),p_{1})]
\nonumber\\&&\times\langle j\nu_j|l\mu_ls\mu_s\rangle \langle s\mu_s|
 s_2 \mu_2 s_3 \mu_3 \rangle Y_{l\mu_l}^*(\hat{\bf k}_{1x})\nonumber\\
&&\times
 D^{s_2}_{\mu_2\nu_2}[R_{cx}(k_{2x})R_{wx}(B_x^{-1}(p_{23}),p_{2})]
 D^{s_3}_{\mu_3\nu_3}[R_{cx}(k_{3x})R_{wx}(B_x^{-1}(p_{23}),p_{3})]
\nonumber\\&&\times
 D^{s_3}_{\nu_3\nu_3'}[R_{wx}(B_x(p_{31}'),k_{3x}')R_{xc}(k_{3x}')]
 D^{s_1}_{\nu_1\nu_1'}[R_{wx}(B_x(p_{31}'),k_{1x}')R_{xc}(k_{1x}')]
\nonumber\\&&\times
 Y_{l'\mu_l'}(\hat{\bf k}_{2x}') \langle s_3\nu_3's_1\nu_1'|s'\mu_s'
 \rangle \langle l'\mu_l's'\mu_s'|j'\nu_j'\rangle \nonumber\\&&
\times D^{j'}_{\nu_j'\mu_j'}[R_{wx}(B_x(P'),q_{31x}')R_{xc}(q_{31x}')]
 D^{s_2}_{\nu_2\mu_2'}[R_{wx}(B_x(P'),q_{2x}')R_{xc}(q_{2x}')]
 \nonumber\\
&&\times Y_{L'\mu_L'}(\hat{\bf q}_{2x}') \langle j'\mu_j' s_2\mu_2' |
 S'\mu_S'\rangle \langle L'\mu_L'S'\mu_S' | J'\mu' \rangle~.
\label{eq:BR}
\end{eqnarray}
Following Balian and Br\'ezin\cite{Balian}, symmetry principles are
used to evaluate this matrix element.  To facilitate the evaluation of
the matrix element $\langle1|2'\rangle$ note that
\begin{eqnarray}
\lefteqn{\langle {\bf P} \mu ; q_1 J L_1 S_1 k_1 j_1 l_1 s_1 |
 {\bf P}' \mu' ; q_2' J' L_2' S_2' k_2' j_2' l_2' s_2' \rangle}
\nonumber\\&& = \langle {\bf P} \mu ; q_1 J L_1 S_1 k_1 j_1 l_1 s_1
 | U^{\dagger}(\Lambda ,a) U(\Lambda , a) |
 {\bf P}' \mu' ; q_2' J' L_2' S_2' k_2' j_2' l_2' s_2' \rangle
\label{eq:BS}
\end{eqnarray}
for any $\Lambda$ and $a$.  The kinematic quantities are
\begin{equation}
{\bf k}_1 = {\bf k}_{2x} = -{\bf k}_{3x}, \qquad
{\bf k}_2' = {\bf k}_{3x}' = -{\bf k}_{1x}', \qquad
{\bf q}_{1x} = -{\bf q}_{23x}, \qquad
{\bf q}_{2x}' = -{\bf q}_{31x}'
\label{eq:BT}
\end{equation}
with
\begin{equation}
k_{2x} = B_x^{-1} (p_{23})p_2, \qquad
k_{3x} = B_x^{-1} (p_{23})p_3, \qquad
k_{3x}' = B_x^{-1} (p_{31}')p_3', \qquad
k_{1x}' = B_x^{-1} (p_{31}')p_1'
\label{eq:BU}
\end{equation}
and
\begin{equation}
q_{1x} = B_x^{-1} (P)p_1, \qquad
q_{23x} = B_x^{-1} (P)p_{23}, \qquad
q_{2x}' = B_x^{-1} (P)p_2', \qquad
q_{31x}' = B_x^{-1} (P)p'_{31}.
\label{eq:BV}
\end{equation}
Evaluating the right hand side of Eq.~(\ref{eq:BS}) gives
\begin{eqnarray}
\langle 1 | 2' \rangle &=& e^{i \Lambda (P'-P)\cdot a} \Biggl|
 {\omega_{M_0}({\bf P}_\Lambda)\over\omega_{M_0}({\bf P})}
 {\omega_{M_0'}({\bf P}'_\Lambda)\over\omega_{M_0'}({\bf P}')}
 \Biggr|^{1/2} D^{J*}_{\nu\mu}[R_{wx}(\Lambda ,P)]
 D^{J}_{\nu'\mu'}[R_{wx}(\Lambda,P')] \nonumber\\
&&\times\langle {\bf P}_{\Lambda}\nu ; q_1 J L_1 S_1 k_1 j_1 l_1 s_1 |
 {\bf P}'_{\Lambda}\nu';q_2' J' L_2' S_2' k_2' j_2' l_2' s_2'\rangle~.
\label{eq:BW}
\end{eqnarray}
To deduce the implications of Eq.~(\ref{eq:BW}) consider different
choices of $\Lambda$ and $a$.  For $\Lambda=I$ and arbitrary $a$ the
relations~(\ref{eq:BW}) cannot be satisfied unless $P=P'$.  For
$\Lambda = B_x (P)$, $P=P'=P_0 =(M_0,0,0,0)$, and $a=0$ it follows
that
\begin{eqnarray}
\lefteqn{\langle {\bf P}\mu ; q_1 J L_1 S_1 k_1 j_1 l_1 s_1 | {\bf P}'
 \mu' ; q_2' J' L_2' S_2' k_2' j_2' l_2' s_2' \rangle} \nonumber\\
&=&\Biggl| {{M_0}\over\omega_{M_0}({\bf P})}
 {{M_0'}\over\omega_{M_0'}({\bf P}')} \Biggr|^{1/2}
 \langle {\bf P}_0 \mu ; q_1 J L_1 S_1 k_1 j_1 l_1 s_1  | {\bf P}'_0
 \mu' ; q_2' J' L_2' S_2' k_2' j_2' l_2' s_2' \rangle
\label{eq:BX}
\end{eqnarray}
which shows that Eq.~(\ref{eq:BW}) can be expressed as
$\delta({\bf P}-{\bf P}')$ multiplied by a quantity independent of
${\bf P}$.  For the case that $\Lambda=R$ is a rotation, $P=P_0=RP_0$,
and $a=0$ it follows that
\begin{eqnarray}
\lefteqn{\langle {\bf P}_0 \mu ; q_1 J L_1 S_1 k_1 j_1 l_1 s_1 |
 {\bf P}'_0 \mu' ; q_2' J' L_2' S_2' k_2' j_2' l_2' s_2' \rangle}
\nonumber\\&=& \langle {\bf P}_0 \nu ; q_1 J L_1 S_1 k_1 j_1 l_1 s_1 |
 {\bf P}'_0 \nu' ; q_2' J' L_2' S_2' k_2' j_2' l_2' s_2' \rangle
\nonumber\\&&\times
 D^{J*}_{\nu\mu}[R_{wx}(R,P_0)]D^{J}_{\nu'\mu'}[R_{wx}(R,P_0)]~.
\label{eq:BY}
\end{eqnarray}
The left hand side of Eq.~(\ref{eq:BY}) is independent of $R$.  Since
the rest boosts $B_x(P_0)$ are rotations (normally chosen to be the
identity), $R_{wx}(R,P_0)$ is a representation of $SU(2)$ which can be
parameterized by elements of $SU(2)$ and integrated over the group.
Since the Haar measure\cite{Pont} is normalized to unity the integral
is still equal to the left side of Eq.~(\ref{eq:BY}).  The integral
over the $D$ functions can be done explicitly using\cite{Gott}
\begin{equation}
\int dR \, D^{J*}_{\nu\mu}[R_{wx}(R,P_0)]
 D^{J'}_{\nu'\mu'}[R_{wx}(R,P_0)] = {1\over2J+1}
 \delta_{JJ'} \delta_{\nu\nu'} \delta_{\mu\mu'}~,
\label{eq:BZ}
\end{equation}
with the result that
\begin{eqnarray}
\lefteqn{\langle {\bf P}_0 \mu ; q_1 J L_1 S_1 k_1 j_1 l_1 s_1  |
 {\bf P}'_0 \mu' ; J' q_2' L_2' S_2' k_2' j_2' l_2' s_2' \rangle}
\nonumber\\&& = \delta_{JJ'} \delta_{\mu\mu'} {1\over2J+1}
 \langle {\bf P}_0 \nu ; q_1 J L_1 S_1 k_1 j_1 l_1 s_1  |
 {\bf P}'_0 \nu ; J' q_2' L_2' S_2' k_2' j_2' l_2' s_2' \rangle~.
\label{eq:CA}
\end{eqnarray}

The general form of the matrix element is obtained by incorporating
all of the consequences of Eq.~(\ref{eq:BW}):
\begin{eqnarray}
\lefteqn{\langle {\bf P} \mu ; q_1 J L_1 S_1 k_1 j_1 l_1 s_1  |
 {\bf P}' \mu' ; q_2' J' L_2' S_2' k_2' j_2' l_2' s_2' \rangle}
\nonumber\\&=& \delta_{JJ'} \delta_{\mu\mu'} \delta({\bf P}-{\bf P}')
 \delta[M_0({\bf q}_1,{\bf k}_1)-M_0({\bf q}_2',{\bf k}_2')]
\nonumber\\&&\times
 A_J(q_1 L_1 S_1 k_1 j_1 l_1 s_1; q_2' L_2' S_2' k_2' j_2' l_2' s_2')
\label{eq:CB}
\end{eqnarray}
where the amplitude $A_J(1;2)$ is invariant under the non-interacting
representation of the Poincar\'e group.  From the
definition~(\ref{eq:CB}) the amplitude $A_J(1;2)$ can be computed from
the matrix element $\langle1|2'\rangle$ in Eq.~(\ref{eq:BR}),
\begin{eqnarray}
\lefteqn{\delta(M-M') A_J(1;2')} \nonumber\\
&&={1\over2J+1}\int d^3P\langle{\bf P}\mu;q_1 J L_1 S_1 k_1 j_1 l_1
 s_1 | {\bf P}_0 \mu ; q_2' J L_2' S_2' k_2' j_2' l_2' s_2' \rangle~.
\label{eq:CC}
\end{eqnarray}
Since $A_J(1;2')$ is invariant, the computation of this quantity is
facilitated by evaluating this expression for ${P}'=P_0$.

The last step used by Balian and Br\'ezin is to exploit the rotational
invariance.  To do this in the relativistic case let the total four
momentum be $P_0=RP_0$ and evaluate
\begin{eqnarray}
\lefteqn{\langle {\bf P}_0 \mu ; q_1 J L_1 S_1 k_1 j_1 l_1 s_1 |
 {\bf P}_0' \mu ; q_2' J' L_2' S_2' k_2' j_2' l_2' s_2' \rangle}
\nonumber\\&=&\int\langle{\bf P}_0\mu; q_1 J L_1 S_1 k_1 j_1 l_1 s_1 |
 U^{\dagger}(R)U(R) | {\bf p}_1 \mu_1 {\bf p}_2 \mu_2 {\bf p}_3 \mu_3
 \rangle d^3p_1 d^3p_2 d^3p_3 \nonumber\\
&&\times \langle {\bf p}_1 \mu_1 {\bf p}_2 \mu_2 {\bf p}_3 \mu_3 |
 U^{\dagger}(R)U(R) | {\bf P}_0' \mu ;
 q_2' J' L_2' S_2' k_2' j_2' l_2' s_2' \rangle \nonumber\\
&=& \int D^{J*}_{\nu \mu}(R) \langle {\bf P}_0 \nu ; q_1 J L_1 S_1
 k_1 j_1 l_1 s_1 | R{\bf p}_1 \mu_1 R{\bf p}_2 \mu_2 R{\bf p}_3 \mu_3
 \rangle d^3p_1 d^3 p_2 d^3 p_3 \nonumber\\
&&\times \langle R{\bf p}_1 \mu_1 R{\bf p}_2 \mu_2 R{\bf p}_3 \mu_3 |
 {\bf P}_0' \nu' ; q_2' J' L_2' S_2' k_2' j_2' l_2' s_2' \rangle
 D^J_{\nu'\mu}(R) \nonumber\\
&=& \int \langle {\bf P}_0 \nu ; q_1 J L_1 S_1 k_1 j_1 l_1 s_1 |
 R{\bf p}_1 \mu_1 R{\bf p}_2 \mu_2 R{\bf p}_3 \mu_3 \rangle
 d^3p_1 d^3p_2 d^3p_3 \nonumber\\
&&\times \langle R{\bf p}_1 \mu_1 R{\bf p}_2 \mu_2 R{\bf p}_3 \mu_3 |
 {\bf P}_0' \nu ; q_2' J' L_2' S_2' k_2' j_2' l_2' s_2' \rangle~.
\label{eq:CD}
\end{eqnarray}
This equation means that the matrix element $\langle1|2'\rangle$
averaged over the magnetic quantum numbers is invariant under
simultaneous rotations of all of the single particle momenta that
appear as intermediate states.  The rotated ${\bf q}_i$'s and
${\bf k}_i$'s obtained by rotating the single particle momenta are
(for $P=P_0$)
\begin{equation}
{\bf q}_{Ri} = R_{wx}(R,P_0) {\bf q}_i~, \qquad
{\bf k}_{Ri} = R_{wx}(R,q_{jk}) {\bf k}_i~.
\label{eq:CE}
\end{equation}

To apply the above result to the computation of Eq.~(\ref{eq:BR}),
note that there are nine delta functions and eight variables of
integration.  Three of the delta functions lead to the overall
momentum conserving delta function, leaving six delta functions.
Since the ${\bf k}$ do not transform simply under rotations, it is
practical to use four of the remaining six delta functions to perform
the integrals over $\hat{{\bf k}}_1$ and $\hat{{\bf k}}_2$.  Two delta
functions remain.  One factors out of the expression, giving the
conservation of the invariant mass.  The other can be used to fix the
angle between $\hat{{\bf q}}_1$ and $\hat{{\bf q}}_2$.  Three
integrals over the two unit vectors $\hat{{\bf q}}_1$ and
$\hat{{\bf q}}_2$ with fixed $\hat{{\bf q}}_1\cdot\hat{{\bf q}}_2$
remain.  If the matrix element is averaged over the magnetic quantum
numbers then the resulting integrand is necessarily independent of the
remaining three variables of integration.  The integral is $8\pi^2$
multiplied by the integrand.  The result, after computing all of the
Jacobians needed to convert the delta functions to the desired form,
is:
\begin{eqnarray}
\langle 1 | 2' \rangle &=& \delta({\bf P}-{\bf P}') \delta(M-M')
 {8\pi^2\over2J+1} {1\over|{\bf k}_{1x}| |{\bf k}_{2x}'|
 |{\bf q}_{1x}| |{\bf q}_{2x}'|} \nonumber\\
&&\times \Biggl| {m_{23}^{3}m'^3_{31}\over\omega_{m_2}({\bf k}_{1x})
 \omega_{m_3}({\bf k}_{1x}) \omega_{m_1}({\bf q}_{1x})
 \omega_{m_{23}}({\bf q}_{1x}) \omega_{m_3}({\bf k}_{2x}')
 \omega_{m_1}({\bf k}_{2x}') \omega_{m_2}({\bf q}_{2x}')
 \omega_{m_{31}}({\bf q}_{2x}')} \Biggr|^{1 /2} \nonumber\\
&&\times \langle J \mu | L \mu_L S \mu_S \rangle \langle S \mu_S |
 j \mu_j s_1 \mu_1 \rangle Y_{L\mu_L}^*(\hat{{\bf q}}_{1x})\nonumber\\
&&\times
 D^{j}_{\mu_j\nu_j}[R_{cx}(q_{23x})R_{wx}(B_x^{-1}(P_0),q_{23x})]
 D^{s_1}_{\mu_1\nu_1}[R_{cx}(q_{1x})R_{wx}(B_x^{-1}(P_0),q_{1x})]
\nonumber\\&&\times
 \langle j \nu_j | s \mu_s l \mu_l \rangle \langle s \mu_s | s_2 \mu_2
 s_3 \mu_3 \rangle Y_{l\mu_l}^*(\hat{{\bf k}}_{1x}) \nonumber\\
&&\times
 D^{s_2}_{\mu_2\nu_2}[R_{cx}(k_{2x})R_{wx}(B_x^{-1}(q_{23x}),k_{2x})]
 D^{s_3}_{\mu_3\nu_3}[R_{cx}(k_{3x})R_{wx}(B_x^{-1}(q_{23x}),k_{3x})]
\nonumber\\ &&\times
 D^{s_3}_{\nu_3\nu_3'}[R_{wx}(B_x(q_{31x}'),k_{3x}')R_{xc}(k_{3x}')]
 D^{s_1}_{\nu_1\nu_1'}[R_{wx}(B_x(q_{31x}),k_{1x}')R_{xc}(k_{1x}')]
\nonumber\\ &&\times
 Y_{l'\mu_l'}(\hat{{\bf k}}_{2x}') \langle s_3 \nu_3' s_1 \nu_1' |
 s' \mu_s' \rangle \langle l' \mu_l' s' \mu_s' | j' \nu_j' \rangle
\nonumber\\ &&\times
 D^{j'}_{\nu_j'\mu_j'}[R_{wx}(B_x(P_0'),q_{31x})R_{xc}(q_{31x}')]
 D^{s_2}_{\nu_2\mu_2'}[R_{wx}(B_x(P_0'),q_{2x}')R_{xc}(q_{2x}')]
\nonumber\\ &&\times
 Y_{L'\mu_L'}(\hat{{\bf q}}_{2x}') \langle j' \mu_j' s_2 \mu_2' |
 S' \mu_S' \rangle \langle L' \mu_L' S' \mu_S' | J \mu \rangle~.
\label{eq:CF}
\end{eqnarray}
The invariant part of Eq.~(\ref{eq:CF}) has been evaluated with
$P=P_0$, which implies the replacement of all of the $p_i$'s by the
corresponding $q_i$'s.  In the case of the front form or point form
the spin must be chosen accordingly (front-from spin for front-form
interactions, canonical spin for the point-form interactions) and the
delta functions in Eq.~(\ref{eq:CF}) are replaced by
\begin{equation}
\delta({\bf P}-{\bf P}') \delta(M-M') \to \delta(P^+-P^{+\prime})
 \delta(P_1-P_1') \delta(P_2-P_2') \delta(M-M')
\label{eq:CG}
\end{equation}
in the front form, and by
\begin{equation}
\delta({\bf P}-{\bf P}') \delta(M-M') \to \delta({\bf V}-{\bf V}')
 {\delta(M-M') \over M}
\label{eq:CH}
\end{equation}
in the point form, where $\{P^+,P_1,P_2\}$ are the front-form
components of the four-momentum and ${\bf V}={\bf P}/M$ are the
independent components of the four-velocity.

Note that if the magnitude of all of the ${\bf k}$'s and ${\bf q}$'s
appearing in both the rotation matrices and the Jacobian factor in
Eq.~(\ref{eq:CF}) are set equal to zero this expression has the
non-relativistic expression (\ref{eq:AN}) as a limit provided the rest
boosts are chosen to be the identity.  Specifically the $D$ functions
all become the identity and the factor inside the $|\cdots|^{1/2}$
becomes $m_{13}m_{23}/m_1m_2m_3$.  The resulting expression is
identical with Eq.~(\ref{eq:AN}).

The expression~(\ref{eq:CF}) explicitly involves the four unit vectors
$\hat{{\bf q}}_1$, $\hat{{\bf q}}_2'$, $\hat{{\bf k}}_1$, and
$\hat{{\bf k}}_2'$.  The $\hat{{\bf q}}_i'$ can be evaluated in any
geometry, subject to the constraint that the angle between the two
unit vectors is fixed by kinematic considerations.  These choices fix
the quantities $\hat{{\bf k}}_i$.  Different choices of the geometry
used to evaluate the $\hat{{\bf q}}_i'$ can lead to additional
simplifications in the evaluation of Eq.~(\ref{eq:CF}), although in
the relativistic case the choice of best geometry depends to some
extent on the choice of spin.  If the boosts satisfy $B_x(P_0)=I$ then
the rest Wigner rotations $R_{wx}[B_x^{-1}(P_0)q)]$ in
Eq.~(\ref{eq:CF}) can be replaced by the identity.  For canonical spin
there are no generalized Melosh rotations, while for the front-from
spin the Wigner rotations of front-form boosts are the identity,
$R_{wf}(B_f(p),q)=I$.  All of these properties lead to further
simplifications of Eq.~(\ref{eq:CF}).

Each of the choices of spin and continuous variables in a
Bakamjian-Thomas model implies a choice of representation.  Although
it is tempting to formulate a model using a choice that leads to the
simplest Racah coefficient, any choice has implications for the
structure of the representation of other operators, such as
electromagnetic and weak current operators.  The interacting system
has interactions in different Poincar\'e generators for different
choices of the Racah coefficients.  For instance, the Racah
coefficients leading to an instant-form dynamics imply that the
infinitesimal generators of rotationless Lorentz transformation
contain interactions, while Racah coefficients appropriate for a
point-form dynamics imply interactions in the momentum operators.  For
more general choices of Racah coefficient there can be interactions in
any number of Poincar\'e generators. For an operator, such as an
electromagnetic current operator, that is well approximated in an
impulse approximation in one representation may require large two-body
contributions in another representation.  Thus, other considerations
may be important in choosing a representation.

As in the non-relativistic case, it is useful to replace the delta
function in the kinematic mass by a delta function that expresses the
invariant mass explicitly in terms of the relative momenta,
\begin{equation}
\delta(M-M') \to \delta(M(k_1,q_1)-M(k_2',q_2'))~,
\label{eq:CHAA}
\end{equation}
where in computing this delta function it is important to note that
only three of the variable can be considered independent.  This
replacement allow one to do the integral over one of the relative
momenta.

The analysis above shows that the methods suggested by Balian and
Br\'ezin is applicable to relativistic three-body equations of the
Bakamjian-Thomas type.  The relativistic expressions for the
recoupling coefficients were shown reduce to the non-relativistic ones
in the non-relativistic limit.

\end{document}